# On the buoyancy force and the metacentre

Jacques Mégel and Janis Kliava

UFR de Physique, Université de Bordeaux 1

351 cours de la Libération,

33405 Talence cedex, France

## Abstract

We address the point of application A of the buoyancy force (also known as the Archimedes force) by using two different definitions of the point of application of a force, derived one from the work-energy relation and another one from the equation of motion. We present a quantitative approach to this issue based on the concept of the hydrostatic energy, considered for a general shape of the immersed cross-section of the floating body. We show that the location of A depends on the type of motion experienced by the body. In particular, in vertical translation, from the work-energy viewpoint, this point is fixed with respect to the centre of gravity G of the body. In contrast, in rolling/pitching motion there is duality in the location of A ; indeed, the work-energy relation implies A to be fixed with respect to the centre of buoyancy C, while from considerations involving the rotational moment it follows that A is located at the metacentre M. We obtain analytical expressions of the location of M for a general shape of the immersed cross-section of the floating body and for an arbitrary angle of heel. We show that three different definitions of M *viz.*, the "geometrical" one, as the centre of curvature of the buoyancy curve, the Bouguer's one, involving the moment of inertia of the plane of flotation, and the "dynamical" one, involving the second derivative of the hydrostatic energy, refer to one and the same special point, and we demonstrate a close relation between the height of M above the line of flotation and the stability of the floating body. Finally, we provide analytical expressions and graphs of the buoyancy, flotation and metacentric curves as functions of the angle of heel, for some particular shapes of the floating bodies, *viz.*, a circular cylinder, a rectangular box, a parabolic and an elliptic cylinder.

## I. Introduction

*"If a solid lighter than a fluid be forcibly immersed in it, the solid will be driven upwards by a force equal to the difference between its weight and the weight of the fluid displaced."* [1, p. 257]

One of the most long-standing – for 22 centuries – and, perhaps, one of the best-known laws of physics is the Archimedes' Law stating that a body immersed in a fluid is subjected to a force equal to the weight of the displaced fluid and acting in opposite direction to the force of gravity. This force is called "buoyancy force" or "Archimedes' force"; we denote it by $\boldsymbol{F}_\mathrm{h}$, the subscript "h" standing for "hydrostatic", and its point of application will be referred to as A.

Clearly, $\boldsymbol{F}_h$ is a resultant of all elementary hydrostatic forces applied to the surface of the immersed body. A force is fully described by (i) its norm, (ii) its direction and (iii) its point of application. Note that the point (iii) is not included in the above definition of the Archimedes' Law. On the other hand, it would be surprising if such a fundamental epistemological question as that of the point of application of the buoyancy force were not at all discussed in the literature.

By definition of the resultant force, the point of application of $\boldsymbol{F}_h$ should be defined by requiring that if we apply such a force, the (macroscopic) behaviour of the body will be the same as that caused by the ensemble of elementary hydrostatic forces. Note, however, that this definition does not guarantee the uniqueness of this point; indeed, one may quite well conceive it to be different in different physical situations.

An analysis of the abundant bibliography concerned with the stability of immersed bodies shows that many authors prefer to elude the question of the exact location of the point of application of the Archimedes' force, $e.g.$, speaking of "the line of action" of this force. [2, p. 73] What is really shown is only that the *line* of application of the Archimedes' force passes through the centre of buoyancy, so that one may guess that A can be located anywhere on this line. If this statement is sufficient in statics, it is certainly not sufficient in dynamics. Indeed, one might argue that a force applied to a *rigid* body, as a sliding vector, obeys to the principle of transmissibility stating that "*the condition of motion of a rigid body remains unchanged if a force F of a given magnitude, direction and sense acts anywhere along the same line of action…*". [3] However, in this relation, an important amendment had been made already more than a century ago, namely, that "*we may imagine a force to be applied at any point in the line of its direction, provided this point be rigidly connected with the first point of application*". [4] Below, we will show that, while a floating body can be considered as rigid, the point of application of $\boldsymbol{F}_h$ is *not* necessarily rigidly connected with it because the displaced fluid is evidently not rigid. Therefore the question of the point of application of this force is quite legitimate and meaningful.

In a quite natural way one is tempted to relate the point of application of $\boldsymbol{F}_h$ to one of the three following remarkable points of a floating body.

(i) The *centre of gravity of the body*, denoted by G (the distinction between the centre of gravity and that of mass/inertia is of no relevance for the present analysis). To our knowledge, such assignment has never been suggested before.

(ii) The *centre of buoyancy* (centre of gravity of the displaced fluid) denoted by C (from *careen*) or by B (from *buoyancy*); we prefer using the former notation. Note that C sometimes is defined as the geometrical centre of the displaced fluid; meanwhile, for a homogeneous fluid (the case considered below) both definitions coincide. Most frequently, the point of application of $\boldsymbol{F}_h$ is assigned to the centre of buoyancy, e.g., see Refs. [5 (p. 58), 6, 7].

(iii) The *metacentre* denoted by M. This point is of a special interest, since, in order to assert the stability of the body against overturn, M, defined in the vicinity of equilibrium, should be located *above* G. We have found only one textbook suggesting, without demonstration, M as a candidate for the point of application of $\boldsymbol{F}_h$, see Ref. [8, footnote p. 26]. On the other hand, in a very interesting while, unfortunately, not easily accessible paper, Herder and Schwab [9] suggest a distinction between "statically equivalent" and "dynamically equivalent" resultant forces. The former are defined for a fixed position of the body (statics) in which case only a *line* and not a *point* of application of the resultant can be defined. The latter are related to the stability of a nominal state with respect to small variations about this state. In this case the point of application of the resultant force is essential, and these authors define it as the "dynamically equivalent" point of application. For the particular case of floating parallelepiped ("shoe-box") Herder and Schwab deduce from considerations based on the hydrostatic energy that the "dynamically equivalent" point of application of the buoyancy force is the metacentre. It has seemed quite tempting to generalize this approach to the general case of the floating body of arbitrary shape, inclined through an arbitrary angle.

The concept of metacentre dates back to Bouguer's *Traité du Navire, de sa construction et de ses mouvemens* (1746) [10]. Using the prefix μετά = "beyond", he had designated a specific point M of a floating body, defined as the intersection of two vertical axes passing through the centre of buoyancy C (the centre of gravity of the displaced fluid) at two slightly different angles of heel. Besides, Bouguer formulated the well-known theorem relating the distance between C and M to the ratio of the moment of inertia of the plane of flotation and the volume of the displaced fluid.

Three years later, Euler in *Scientia navalis* (1749) gave a general criterion of the ship stability, based on the restoring moment: [11] the ship remains stable as far as the couple weight (applied at G) and the buoyancy force (whose line of application passes through C) creates a restoring moment. A change of sign of the latter results in capsizing, and its vanishing in inclined position (at equilibrium it vanishes by definition) corresponds to the overturn angle.

The problem of stability of the floating bodies, which can be traced back to Archimedes himself, see [1, *On floating bodies, Book I,* pp. 253-262; *Book II,* pp. 263-300], has never ceased to interest scientists and engineers [12-16], in particular, in the relation to the metacentre, e.g., see [7, 17-20] and has become an important part of academic studies [2, 5, 6, 8, 21]. Meanwhile, there has been no significant progress in this field since the original Bouguer's Treatise [10] and its reformulation in terms of a "novel geometry" in the Dupin's textbook [17], relating the *metacentric curve, i.e.,* the loci of M, with the *buoyancy curve* formed by the loci of C, the former being the *evolute* of the latter.

The aim of the present work is to elucidate the question of the point of application of the buoyancy force in relation with that of ship stability. We consider two different approaches to the definition of the point of application of a resultant force based, first, on the work-energy relation and, second, on the equation of motion. We obtain appropriate expressions of the hydrostatic energy and the location of the characteristic points for a floating body of rather general shape. On this basis we show that the location of the point of application of the buoyancy force depends on the type of motion experienced by the body. In particular, in pure translation this point is fixed with respect to the centre of gravity G while in rolling or pitching motion it is fixed with respect to the centre of buoyancy C (from the viewpoint of the work-energy relation) or located at the metacentre M (from the viewpoint of the rotational moment).

In the framework of the formalism developed in this work, we present a quantitative analysis of the concept of the metacentre. We calculate the location of the "geometrical", Bouguer's and "dynamical" metacentres for an arbitrary angle of heel and show that all three definitions, in fact, result in one and the same metacentric curve. Finally, in the Appendix we apply the general relationships to determine the location of C and M for several particularly simple shapes of the floating body.

## II. Buoyancy force and hydrostatic energy

Consider a *floating, i.e.*, partially immersed in a fluid, *rigid* body, *e.g.*, a vessel. In such situation, strictly speaking, one should also take into account the atmospheric pressure experienced by the part of the body situated above the waterline. However, as far as the specific density of the air remains much lower than that of the fluid, as we have assumed in the analysis given below, the atmospheric pressure can be neglected. In the formulae given below the surface $S$ and the volume $V$ concern only the submerged part of the body, $V$ being equal to the volume of the displaced fluid.

The hydrostatic force exerted on an element $\mathrm{d}\boldsymbol{S}$ of the immersed surface of the body is $\mathrm{d}\boldsymbol{F}_{\mathrm{h}} = -p\,\mathrm{d}\boldsymbol{S}$ where $p$ is the hydrostatic pressure. The minus sign in this expression is due to the fact that, by convention, $\mathrm{d}\boldsymbol{S}$ is directed outwards a closed surface while the hydrostatic force is directed inwards the body. Integrating $\mathrm{d}\boldsymbol{F}_{\mathrm{h}}$ over the immersed surface and applying the gradient theorem to pass from a surface integral to an integral over the immersed volume $V$ yields the buoyancy force as:

$$\boldsymbol{F}_{\mathrm{h}} = -\oiint_{S} p\,\mathrm{d}\boldsymbol{S} = -\iiint_{V} \nabla p\,\mathrm{d}V \,. \tag{II.1}$$

Eq. (II.1) shows that $\boldsymbol{F}_{\mathrm{h}}$ can also be considered as a resultant of *fictitious* volume forces of density per unit volume of the fluid $\mathrm{d}\boldsymbol{F}_{\mathrm{h}}/\mathrm{d}V = -\nabla p$, deriving from a potential energy (hydrostatic energy) of density $p$. The total hydrostatic energy is calculated as

$$E_{\mathrm{h}} = \iiint_{V} p\,\mathrm{d}V \,. \tag{II.2}$$

From the general relation between a force and the corresponding potential energy it follows that $\boldsymbol{F}_\mathrm{h} = -\nabla E_\mathrm{h}$.

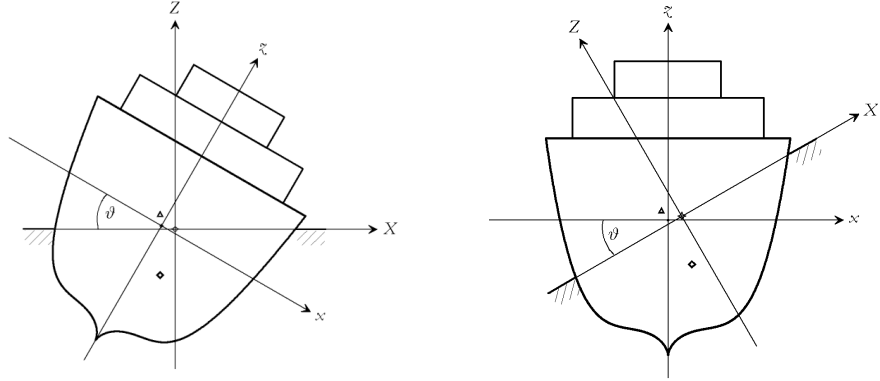

Figure 1. Two different representations of the coordinate systems defined in the text. Left: The floating body turned anti-clockwise with respect to the horizontal line of flotation. Right: The line of flotation turned clockwise with respect with the vertical floating body. The profile of the immersed cross-section is described by the function $3x^4 - x^2 + \frac{1}{2}\sqrt{|x|} - 1$. The diamonds, circles and triangles indicate, respectively, calculated locations of the C, O and M points.

One defines the *plane of flotation* as "the plane in which the body is intersected by the surface of the liquid" [22, p. 67]. The *line of flotation* is defined as the intersection of the plane of flotation with a vertical cross-section of the body; the centre of the former is referred to as the *centre of flotation* and denoted by O.

Below we will consider two types of motion of a partially immersed body: a vertical translation and an oscillation about one of the horizontal axes of a floating body. In most cases a floating vessel is much longer in one of the principal horizontal directions than in the perpendicular one, and the oscillations about the longitudinal and transversal axes are respectively called *rolling* and *pitching*. We will not explicitly consider the pitching; indeed, its characteristics can be readily obtained from those of the rolling motion. The latter will be systematically described in two interrelated coordinate systems, see Figure 1. The main axes of the first one, for brevity referred to as the *Earth frame*, are denoted by majuscules and chosen as follows: the $X$ and $Y$ axes are respectively directed along the width and the length of the body and the $Z$ axis is vertical ascending; the corresponding unit vectors are denoted by $\boldsymbol{u}_X, \boldsymbol{u}_Y, \boldsymbol{u}_Z$. The origin of coordinates is chosen at the *instantaneous* position of the centre of flotation O, and the horizontal $XOY$ plane coincides with the plane of flotation. The second coordinate system, referred to as the *body frame*, is rigidly related to the body. Its axes are denoted by minuscules, with the $y$ axis along the longitudinal axis of the body and the $x$ and $z$ axes turned through an angle $\vartheta$ (angle of heel) with respect to the $X$ and $Z$ axes of the Earth frame; the corresponding unit vectors are denoted by $\boldsymbol{u}_x, \boldsymbol{u}_y, \boldsymbol{u}_z$. The origin of coordinates is chosen at the location of O *at equilibrium*, therefore, at equilibrium both frames coincide and for an arbitrary angle of heel the relation between the respective coordinates is

$$X = (x - x_O) \cos \vartheta + (z - z_O) \sin \vartheta$$
$$Z = -(x - x_O) \sin \vartheta + (z - z_O) \cos \vartheta .$$

(II.3)

We suppose that, as is most often the case, the fluid can be considered as incompressible and homogeneous, so that the hydrostatic pressure depends only on the immersion depth: $p = -\mu_f g Z$ where $\mu_f$ is the specific density of the fluid and $g$ is the acceleration of gravity. Then eqs. (II.1), (II.2) become:

$$\boldsymbol{F}_h = \mu_f g \iiint_V \mathrm{d}V \boldsymbol{u}_Z = \mu_f g V \boldsymbol{u}_Z$$

(II.4)

and

$$E_h = -\mu_f g \iiint_V Z \, \mathrm{d}V .$$

(II.5)

# III. Two definitions of the point of application of a force

By definition of the centre of buoyancy, its coordinates are

$$X_C = \frac{1}{V} \iiint_V X \, \mathrm{d}V ; \quad Y_C = \frac{1}{V} \iiint_V Y \, \mathrm{d}V ; \quad Z_C = \frac{1}{V} \iiint_V Z \, \mathrm{d}V ,$$

(III.1)

therefore, from eqs. (II.5) and (III.1),

$$E_h = -\mu_f g V Z_C = -F_h Z_C.$$

(III.2)

The concept of the hydrostatic energy allows defining the point of application of the buoyancy force from the usual work-energy relation, e.g., see ref. [22]. The elementary work is defined as the scalar product of a force with an elementary displacement of its point of application; on the other hand, for a force deriving from a potential energy it equals the diminution of this energy. In the present case,

$$\delta W_h = \boldsymbol{F}_h \cdot \mathrm{d}\boldsymbol{r}_A = F_h \, \mathrm{d}Z_A = -\mathrm{d}E_h .$$

(III.3)

If the displacement of the point A matches that of a *definite* point of the body, the latter point can be *identified* as the point of application of $\boldsymbol{F}_h$. As far as the above identification is based on the work-energy relation, the corresponding point will be referred to as the "energetical" point of application of the buoyancy force. This definition applies for any type of motion.

Another possibility of defining the point of application of $\boldsymbol{F}_h$ is based on the principle that the rotational moment of a force vanishes in its point of application. This definition is applicable to types of motion where a rotational component is present, in particular, to rolling/pitching. A floating body experiences both the buoyancy force $\boldsymbol{F}_{\mathrm{h}}$ and the force of gravity $\boldsymbol{F}_{\mathrm{g}}$, and its total potential energy is the sum of the hydrostatic and gravitational energies, $E_{\mathrm{total}} = E_{\mathrm{h}} + E_{\mathrm{g}}$. The couple $\boldsymbol{F}_{\mathrm{h}} + \boldsymbol{F}_{\mathrm{g}}$ produces a rotational moment (torque) $\boldsymbol{M}_{\mathrm{total}} = \boldsymbol{r}_{\mathrm{A}} \wedge \boldsymbol{F}_{\mathrm{h}} + \boldsymbol{r}_{\mathrm{G}} \wedge \boldsymbol{F}_{\mathrm{g}}$ where $\boldsymbol{r}_{\mathrm{A}}$ and $\boldsymbol{r}_{\mathrm{G}}$ are position vectors of A and G with respect to an arbitrary coordinate system; obviously, $\boldsymbol{M}_{\mathrm{total}}$ does not depend on the choice of the origin of coordinates.

The strategy usually adopted in mechanics is to separate a general motion of the body into translation of an arbitrary point, chosen inside or outside the body, and rotation about this point. Most often the centre of rotation is chosen at G, in which case $\boldsymbol{r}_{\mathrm{G}} = 0$, but this choice is not compulsory. For a floating body the location of G depends not only of its shape but also of the distribution of weights inside the hull, which changes with the ship loading. Therefore, we have chosen to focus on the buoyancy force and the related hydrostatic energy without systematically calling to mind the force of gravity and its potential energy.

However, the rotational moment of the single force $\boldsymbol{F}_{\mathrm{h}}$ *does* depend on the origin of $\boldsymbol{r}_{\mathrm{A}}$, therefore, in considering *only* the buoyancy force and the hydrostatic energy, the choice of the origin of coordinates is physically meaningful. The famous Euler's theorem [11] states the following.

"***The oscillatory movement of a floating body (rolling or pitching) can be described as a rotation about the centre of flotation O.***"

Choosing O as the origin of $\boldsymbol{r}_{\mathrm{A}}$, the rotational moment of the buoyancy force, $\boldsymbol{M}_{\mathrm{h}}$ is calculated from a variation of $E_{\mathrm{h}}$:

$$\mathrm{d}E_{\mathrm{h}} = -\mathrm{d}\boldsymbol{r}_{\mathrm{A}} \cdot \boldsymbol{F}_{\mathrm{h}} = -\left(\mathrm{d}\boldsymbol{\Theta} \wedge \boldsymbol{r}_{\mathrm{A}}\right) \cdot \boldsymbol{F}_{\mathrm{h}} = -\mathrm{d}\boldsymbol{\Theta} \cdot \left(\boldsymbol{r}_{\mathrm{A}} \wedge \boldsymbol{F}_{\mathrm{h}}\right) = -\mathrm{d}\boldsymbol{\Theta} \cdot \boldsymbol{M}_{\mathrm{h}} \qquad \text{(III.4)}$$

where the elements of the angle vector $\boldsymbol{\Theta} = \left(\varphi, \quad \vartheta, \quad 0\right)$ are angles of rotation about the $X$ and $Y$ axes. In deriving eq. (III.4) we have used the expression of variation of a position vector $\boldsymbol{r}$ in an infinitesimal rotation, $\mathrm{d}\boldsymbol{r} = \mathrm{d}\boldsymbol{\Theta} \wedge \boldsymbol{r}$. From the latter equation one gets $\boldsymbol{M}_{\mathrm{h}} = -\nabla_{\Theta} E_{\mathrm{h}}$ where elements of the gradient $\nabla_{\Theta}$ are derivatives with respect to the angles of rotation. In the particular case of rolling, the only non-vanishing element of $\boldsymbol{M}_{\mathrm{h}}$ is

$$\boldsymbol{M}_{\mathrm{h}} = -\frac{\partial E_{\mathrm{h}}}{\partial \vartheta} \boldsymbol{u}_{Y}. \qquad \text{(III.5)}$$

On the other hand, $\boldsymbol{M}_{\mathrm{h}}$ is the sum of moments of the elementary hydrostatic forces, so, for a homogeneous fluid it can be calculated as

$$\boldsymbol{M}_{\mathrm{h}} = \oiint_{S} \boldsymbol{r} \wedge \mathrm{d}\boldsymbol{F}_{\mathrm{h}} = \mu_{\mathrm{f}} g \oiint_{S} Z \boldsymbol{r} \wedge \mathrm{d}\boldsymbol{S}. \qquad \text{(III.6)}$$

Applying the curl theorem and taking into account that the curl of the radius vector is identically null, $\nabla \wedge \boldsymbol{r} \equiv \boldsymbol{0}$, we get

$$\boldsymbol{M}_{\mathrm{h}} = -\mu_{\mathrm{f}} g \iiint\limits_{V} \nabla \wedge (Z\boldsymbol{r})\, \mathrm{d}V = \mu_{\mathrm{f}} g \iiint\limits_{V} \left(-X\boldsymbol{u}_Y + Y\boldsymbol{u}_X\right) \mathrm{d}V \,. \tag{III.7}$$

By definition of the moment of the buoyancy force,

$$\boldsymbol{M}_{\mathrm{h}} = \boldsymbol{r}_{\mathrm{A}} \wedge \boldsymbol{F}_{\mathrm{h}} = F_{\mathrm{h}} \boldsymbol{r}_{\mathrm{A}} \wedge \boldsymbol{u}_Z = \mu_{\mathrm{f}} g V \left(-X_{\mathrm{A}} \boldsymbol{u}_Y + Y_{\mathrm{A}} \boldsymbol{u}_X\right) . \tag{III.8}$$

From eqs. (III.7) and (III.8) we obtain the coordinates of A in the horizontal plane as the first moments of the immersed volume about the $X$ and $Y$ axes. These coordinates coincide with the corresponding coordinates of C, *cf.* eqs. (III.1):

$$X_{\mathrm{A}} = \frac{1}{V} \iiint\limits_{V} X\, \mathrm{d}V \quad ; \quad Y_{\mathrm{A}} = \frac{1}{V} \iiint\limits_{V} Y\, \mathrm{d}V \,. \tag{III.9}$$

$X_{\mathrm{A}}$ and $Y_{\mathrm{A}}$ give the lengths of the lever arms of $\boldsymbol{F}_{\mathrm{h}}$ for rolling and pitching, respectively. Therefore, $\boldsymbol{F}_{\mathrm{h}}$ is applied *along the vertical line passing through* C. Meanwhile, as one might expect, the vertical coordinate $Z_{\mathrm{A}}$ of the point of application of $\boldsymbol{F}_{\mathrm{h}}$ remains undefined.

Interestingly, if we attempt to determine the centre of gravity of a body from vanishing of the corresponding rotational moment *in a given position*, its vertical coordinate $Z_{\mathrm{G}}$ will also remain undefined. Meanwhile, this difficulty is readily overcome by rotating the body through an arbitrary angle about any non-vertical axis. From the viewpoint of such experience, the point of application of the weight can be defined as the "intersection of all vertical lines passing through the centre of gravity for different angular positions of the body". The same procedure can be used to specify $Z_{\mathrm{A}}$, and as will be discussed later, the "intersection of all vertical lines passing through the centre of gravity *of the displaced fluid* (i.e. the centre of buoyancy) for different angular positions of the body", by definition, indicates the metacentre.

Therefore, we need an equation of motion involving $\boldsymbol{M}_{\mathrm{h}}$ and containing $Z_{\mathrm{A}}$. In eq. (III.8) $Z_{\mathrm{A}}$ is eliminated because of the properties of the vector product $\boldsymbol{r}_{\mathrm{A}} \wedge \boldsymbol{F}_{\mathrm{h}}$, so, in a somewhat intuitive way, one may suggest an equation satisfied by the corresponding scalar product $\boldsymbol{r}_{\mathrm{A}} \cdot \boldsymbol{F}_{\mathrm{h}}$. It can be obtained considering the equation of rotation of a solid: $\boldsymbol{I}\ddot{\boldsymbol{\Theta}} = \boldsymbol{M}_{\mathrm{h}}$ where $\boldsymbol{I}$ is the tensor of inertia about O and the elements of the angle *vector* $\ddot{\boldsymbol{\Theta}}$, the second derivative of $\boldsymbol{\Theta}$, are angular accelerations around the $X$ and $Y$ axes. In order to account for a small rotation about the state of the body described by this equation, we take the differentials of its both sides:

$$\boldsymbol{I}\, \mathrm{d}\ddot{\boldsymbol{\Theta}} = \mathrm{d}\boldsymbol{M}_{\mathrm{h}} = \left(\mathrm{d}\boldsymbol{\Theta} \wedge \boldsymbol{r}_{\mathrm{A}}\right) \wedge \boldsymbol{F}_{\mathrm{h}} = \boldsymbol{F}_{\mathrm{h}} \wedge \left(\boldsymbol{r}_{\mathrm{A}} \wedge \mathrm{d}\boldsymbol{\Theta}\right) . \tag{III.10}$$

Developing the double vector product and making use of the fact that $\boldsymbol{F}_{\mathrm{h}}$ is vertical while $\mathrm{d}\boldsymbol{\Theta}$ is horizontal, yields

$$\mathrm{d}\boldsymbol{M}_{\mathrm{h}} = \boldsymbol{r}_{\mathrm{A}}\left(\boldsymbol{F}_{\mathrm{h}}\cdot\mathrm{d}\boldsymbol{\Theta}\right) - \mathrm{d}\boldsymbol{\Theta}\left(\boldsymbol{r}_{\mathrm{A}}\cdot\boldsymbol{F}_{\mathrm{h}}\right) = -\mathrm{d}\boldsymbol{\Theta}\left(\boldsymbol{r}_{\mathrm{A}}\cdot\boldsymbol{F}_{\mathrm{h}}\right). \tag{III.11}$$

In the case of the rolling motion, from eqs. (III.5) and (III.11) we get for the sought-after scalar product:

$$\frac{\partial^2 E_{\mathrm{h}}}{\partial \vartheta^2} = -\frac{\mathrm{d}M_{\mathrm{h}_Y}}{\mathrm{d}\vartheta} = \boldsymbol{r}_{\mathrm{A}}\cdot\boldsymbol{F}_{\mathrm{h}} = Z_{\mathrm{A}}F_{\mathrm{h}}. \tag{III.12}$$

Thus, defining the vertical coordinate of the point of application of $\boldsymbol{F}_{\mathrm{h}}$, we are led to the following theorem:

*The scalar product of the buoyancy force and the position vector of its point of application from the origin at the centre of flotation is given by minus the derivative of the rotational moment with respect to the angle of roll or pitch or by the second derivative of the hydrostatic energy with respect to this angle.*

This result can be considered as a particular case of the general analysis of the "dynamically equivalent" point of application given by Herder and Schwab [9].

Eqs. (III.9) and (III.12) are necessary and sufficient to determine all the three coordinates of the point of application of $\boldsymbol{F}_{\mathrm{h}}$. As this determination is based on dynamical equations involving the rotational moment, this point will be referred to as the "dynamical" point of application of the buoyancy force.

It is easy to show that the second derivative of the total potential energy can be expressed as

$$\frac{\partial^2 E_{\mathrm{total}}}{\partial \vartheta^2} = -\frac{\mathrm{d}M_{\mathrm{total}_Y}}{\mathrm{d}\vartheta} = \left(\boldsymbol{r}_{\mathrm{A}} - \boldsymbol{r}_{\mathrm{G}}\right)\cdot\boldsymbol{F}_{\mathrm{h}} = \left(Z_{\mathrm{A}} - Z_{\mathrm{G}}\right)F_{\mathrm{h}}, \tag{III.13}$$

directly yielding the vertical distance between G and A.

In what follows we present a model calculation of the location of the different points of application of the buoyancy force. The calculations refer to the immersed part of the transversal cross-section of the floating body for brevity denoted by "immersed cross-section", so, the results given below refer to a portion of the body of unit length along its longitudinal axis. The characteristics of the whole body can be readily obtained by weighted integration along the $y/Y$ axis.

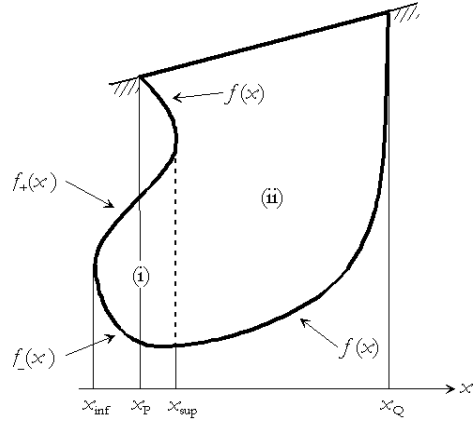

**Figure 2**. Example of an immersed cross-section, (i) and (ii) representing fully and partially immersed pieces.

In the general case the shape of the immersed cross-section can be quite complicated. It is not necessarily supposed to be symmetric with respect to the z axis, e.g., the case of a Venetian gondola; moreover, it can be delimited by a multi-valued function, e.g., see Figure 2. However, it can always be represented as a combination of pieces of two different types:

(i) those fully immersed, delimited from the top and from the bottom by functions $z = f_+(x)$ and $z = f_-(x)$, respectively, and along the horizontal axis by the abscissas $x_{inf}$ and $x_{sup}$;

(ii) those partially immersed, delimited from the top by the line of flotation and from the bottom by a one-value function $z = f(x)$ and along the horizontal axes by the abscissas $x_P$ and $x_Q$.

The expressions of areas and coordinates of the centres of buoyancy for such pieces are given in Appendix I.

## IV. The point of application of the buoyancy force in translation

For definiteness, we consider a vertical ascending motion of the body (a "dry-docking"). This problem is most conveniently treated in the body frame at $\vartheta = 0$. We assume that the body is in equilibrium; therefore, the force of gravity is supposed to be applied along the vertical line passing through C.

First we consider a fully immersed piece of type (i) elevated through a small vertical distance $db$. Its area and hence $\boldsymbol{F}_h$ remain constant, *cf.* eqs. (A.1), and from eq. (III.2) we get

$$dE_h = -dF_h z_C - F_h \, dz_C = -F_h \, dz_C. \tag{IV.1}$$

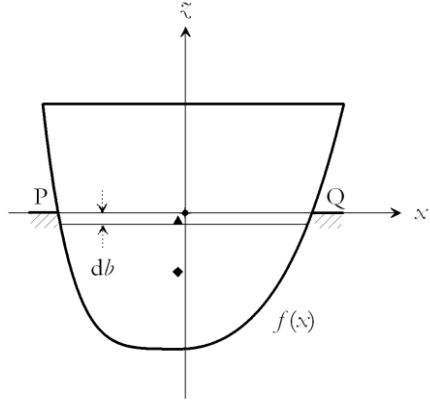

**Figure 3.** Immersed cross-section delimited by the single-valued function $0.9x^6 + 0.9x^4 - 1$ for $x < 0$ and $0.36x^4 + 0.75x^2 - 1$ for $x > 0$. $\mathrm{d}b$ is a small vertical translation. The diamond and circle and triangle indicate, respectively, the locations of C, O and M.

From comparison with the work-energy relation, *cf.* eq. (III.3), it follows that

$$\mathrm{d}z_{\mathrm{A}} = \mathrm{d}z_{\mathrm{C}},\qquad\qquad(\text{IV.2})$$

therefore, the point of application of $\boldsymbol{F}_{\mathrm{h}}$ is rigidly connected with C. Meanwhile, in this case C is rigidly connected with the immersed piece itself, so, A can be assigned to *any* point located on the vertical line through C. One can see that for a fully immersed piece only the line and not the point of application of $\boldsymbol{F}_{\mathrm{h}}$ can be defined.

Next, we treat a partially immersed piece of type (ii), see Figure 3. In the starting position the line of flotation is located at $z_{\mathrm{O}} = 0$, so, the buoyancy force is, *cf.* eq. (II.4),

$$F_{\mathrm{h}} = \mu_{\mathrm{f}}\,g\int\limits_{x_{\mathrm{P}}}^{x_{\mathrm{Q}}}\int\limits_{f}^{0}\mathrm{d}z\,\mathrm{d}x = -\mu_{\mathrm{f}}\,g\int\limits_{x_{\mathrm{P}}}^{x_{\mathrm{Q}}} f\,\mathrm{d}x\,.\qquad\qquad(\text{IV.3})$$

and the hydrostatic energy is, *cf.* eq. (II.5),

$$E_{\mathrm{h}} = -\mu_{\mathrm{f}}\,g\int\limits_{x_{\mathrm{P}}}^{x_{\mathrm{Q}}}\int\limits_{f}^{0} z\,\mathrm{d}z\,\mathrm{d}x = \tfrac{1}{2}\mu_{\mathrm{f}}\,g\int\limits_{x_{\mathrm{P}}}^{x_{\mathrm{Q}}} f^2\,\mathrm{d}x\,.\qquad\qquad(\text{IV.4})$$

When the piece is elevated through $\mathrm{d}b > 0$, the line of flotation moves down to $z = -\mathrm{d}b$, so that the change in $F_{\mathrm{h}}$ is

$$\mathrm{d}F_{\mathrm{h}} = \mu_{\mathrm{f}}\,g\int\limits_{x_{\mathrm{P}}}^{x_{\mathrm{Q}}}\int\limits_{0}^{-\mathrm{d}b}\mathrm{d}z\,\mathrm{d}x = -\mu_{\mathrm{f}}\,gL\,\mathrm{d}b\qquad\qquad(\text{IV.5})$$

where $L = x_{\mathrm{Q}} - x_{\mathrm{P}}$ is the length of the line of flotation for $\vartheta = 0$. The corresponding change in $E_{\mathrm{h}}$ in linear approximation is, *cf.* eq. (II.5),

$$\mathrm{d}E_\mathrm{h} = -\mu_\mathrm{f}g\int\limits_{x_\mathrm{P}}^{x_\mathrm{Q}}\left(\int\limits_f^{-\mathrm{d}b}(z+\mathrm{d}b)-\int\limits_f^0 z\right)\mathrm{d}z\,\mathrm{d}x \approx \mu_\mathrm{f}g\int\limits_{x_\mathrm{P}}^{x_\mathrm{Q}}f\,\mathrm{d}x\,\mathrm{d}b\,. \qquad (\mathrm{IV.6})$$

From eqs. (IV.3) and (IV.6) one gets:

$$\mathrm{d}E_\mathrm{h} = -F_\mathrm{h}\,\mathrm{d}b\,, \qquad (\mathrm{IV.7})$$

so, a comparison with eq. (III.3) shows that

$$\mathrm{d}z_\mathrm{A} = \mathrm{d}b\,. \qquad (\mathrm{IV.8})$$

As in the case of a fully immersed piece, A is rigidly connected with the partially immersed piece in translational motion.

For an immersed cross-section of a general shape the coordinates of A are weighted sums of the corresponding coordinates of separate pieces, therefore A remains rigidly connected with such a piece. As G is rigidly connected with the immersed body and, in translation, G is located on the axis of application of $\boldsymbol{F}_\mathrm{h}$, one can consider that $\boldsymbol{F}_\mathrm{h}$ *is* applied at G. This assignment being based on the work-energy relation, the corresponding point of application of $\boldsymbol{F}_\mathrm{h}$ will be referred to as an "energetical" one.

On the other hand, C is *not* rigidly connected with a partially immersed piece; therefore, in this case it would be an error to consider that $\boldsymbol{F}_\mathrm{h}$ is applied at C. Indeed, from eqs. (III.2) and (IV.7),

$$\mathrm{d}E_\mathrm{h} = -\mathrm{d}F_\mathrm{h}z_\mathrm{C} - F_\mathrm{h}\,\mathrm{d}z_\mathrm{C} = -F_\mathrm{h}\,\mathrm{d}b\,, \qquad (\mathrm{IV.9})$$

and substituting $F_\mathrm{h}$, $E_\mathrm{h}$, $\mathrm{d}F_\mathrm{h}$ and $\mathrm{d}E_\mathrm{h}$ from the respective expressions, one gets

$$\mathrm{d}z_\mathrm{C} = \left[1 - \tfrac{1}{2}L\int\limits_{x_\mathrm{P}}^{x_\mathrm{Q}}f^2\,\mathrm{d}x \left/ \left(\int\limits_{x_\mathrm{P}}^{x_\mathrm{Q}}f\,\mathrm{d}x\right)^2\right.\right]\mathrm{d}b\,. \qquad (\mathrm{IV.10})$$

Obviously, only for a fully immersed body, when $L = 0$, one gets $\mathrm{d}z_\mathrm{C} = \mathrm{d}b$. In other cases the relation between the displacements of the body and of its centre of buoyancy can be very different.

The difference between $\mathrm{d}z_\mathrm{A}$ and $\mathrm{d}z_\mathrm{C}$ is particularly obvious for a partially immersed parallelepiped, $z = f(x) = const = -b$ for $x \in \,]x_\mathrm{P}, x_\mathrm{Q}[$, see Figure 4. In this case $\int_{x_\mathrm{P}}^{x_\mathrm{Q}}f\,\mathrm{d}x = -bL$ and $\int_{x_\mathrm{P}}^{x_\mathrm{Q}}f^2\,\mathrm{d}x = b^2L$, so that for a finite displacement $\Delta b$ eq. (IV.10) yields

$$\Delta z_\mathrm{C} = \tfrac{1}{2}\Delta b\,. \qquad (\mathrm{IV.11})$$

Thus, the displacement of C is only *half* of that of G (and of the point of application of the buoyancy force).

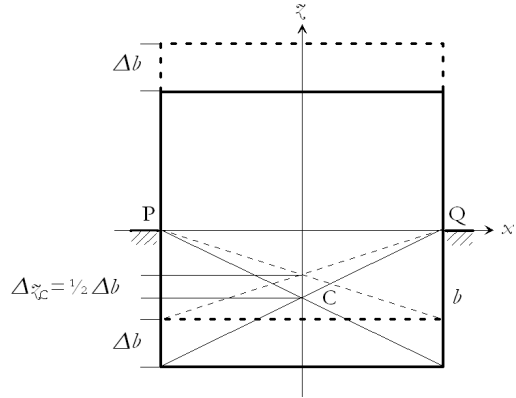

**Figure 4.** Square immersed to a depth $b$. $\Delta_b$ and $\Delta z_C$ are, respectively, vertical translations of the square and of C.

It can be easily shown that if the contour of the immersed cross-section $f(x)$ is a single-valued function, $\mathrm{d}z_C$ is always smaller than or equal to $\frac{1}{2}\mathrm{d}b$. Indeed, from eq. (IV.10) one has

$$\left[\int_{x_P}^{x_Q} f(x)\mathrm{d}x\right]^2 \leq L\int_{x_P}^{x_Q} f^2(x)\mathrm{d}x. \tag{IV.12}$$

In the integral form this inequality was first published on p. 4 of the paper by Bouniakowsky [23], but in can be directly deduced from the Cauchy-Schwarz inequality valid for any square-integrable functions. One can see that $\mathrm{d}z_C \leq \frac{1}{2}\mathrm{d}b$, the limiting case $\mathrm{d}z_C = \frac{1}{2}\mathrm{d}b$ occurring for the rectangle, as discussed before.

In the general case the relation between $\mathrm{d}z_C$ and $\mathrm{d}b$ can be very different. For instance, consider the immersed cross-section shown in Figure 5, looking like the bulb-shape keel of certain yachts participating in the America's cup regatta [24]. Surprisingly, in this case one obtains $\mathrm{d}z_C \approx -\mathrm{d}b$, so, when the body is *elevated* the centre of buoyancy is *lowered* over the same distance !

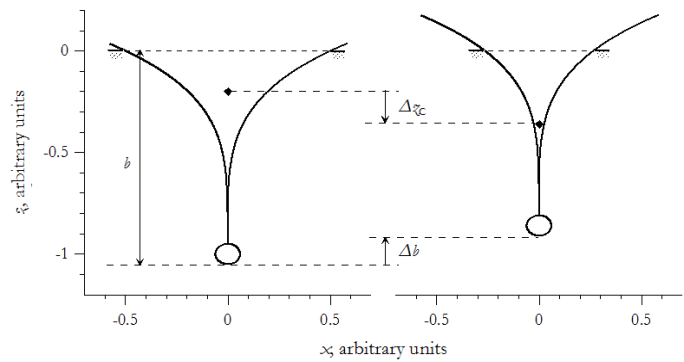

**Figure 5.** Cross-section of a bulb-shape keel. The profile of the keel and the bulb are described, respectively, by $\frac{19}{20}\left[\left(2|x|\right)^{3/4}-1\right]$ and $-1\pm\frac{5}{6}\sqrt{0.06^2-x^2}$. The diamonds indicate the positions of C.

# V. Duality of the point of application of the buoyancy force in rolling/pitching motion

Now we examine the location of the point of application in rolling/pitching. In the general case, in this type of motion a rotation through an angle $\vartheta$ is combined with a vertical translation of the centre of buoyancy, see Figure 6.

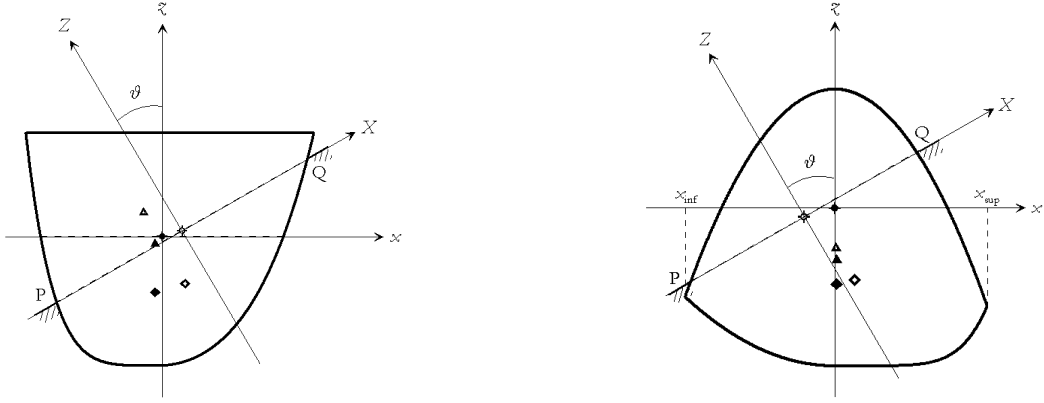

Figure 6. Two immersed cross-sections turned through an angle $\vartheta$. Left: that of Figure 3; right: the one delimited by the functions $f_+(x) = \frac{1}{4} - \frac{3}{2}x^2$ (upper branch) and $f_-(x) = \frac{1}{2}x^2 - 1$ for $x < 0$ and $f_-(x) = \frac{1}{2}x^6 - 1$ for $x > 0$ (lower branch). The full and empty diamonds, circles and triangles indicate, respectively, the locations of C, O and M before and after the rotation.

At an arbitrary angle of heel the line of flotation is described by the equations

$$\frac{z - z_O}{x - x_O} = \frac{z_Q - z_P}{x_Q - x_P} = \tan\vartheta \qquad (V.1)$$

where the coordinates of O are

$$\begin{aligned} x_O &= \tfrac{1}{2}\big(x_P + x_Q\big) \\ z_O &= \tfrac{1}{2}\big(z_P + z_Q\big) \end{aligned} \qquad (V.2)$$

The curve described by O in the course of the rolling/pitching motion is called *flotation curve*.

The area of the immersed cross-section and the coordinates of the centre of buoyancy are given by eqs. (A.1), (A.3) and (A.4).

## V. 1. Energetical considerations

Denoting by $d_C^{\text{flotation}}$ the depth of C below the line of flotation, $E_h$ in the body frame is expressed as

$$E_{\text{h}} = F_{\text{h}} d_{\text{C}}^{\text{flotation}} = F_{\text{h}} \left[ \left( x_{\text{C}} - x_{\text{O}} \right) \sin \vartheta - \left( z_{\text{C}} - z_{\text{O}} \right) \cos \vartheta \right] \tag{V.3}$$

and in the Earth frame (reminding that by definition of the Earth frame, $Z_{\text{O}} = 0$) *cf.* eq. (II.3),

$$E_{\text{h}} = -F_{\text{h}} Z_{\text{C}} . \tag{V.4}$$

In order to calculate the derivative of $E_{\text{h}}$ with respect to $\vartheta$, we consider a small variation of the angle of roll and make use of the fact that the area of the immersed cross-section remains constant in the course of the rolling/pitching motion:

$$\frac{\text{d}\mathscr{A}}{\text{d}\vartheta} \equiv \mathscr{A}' = 0 \tag{V.5}$$

(here and below the primes denote derivatives with respect to $\vartheta$). Substituting $\mathscr{A} = \mathscr{A}_{\text{i}} + \mathscr{A}_{\text{ii}}$ from eqs. (A.1) and applying the Leibniz rule to calculate the derivatives of the parameter-depending integrals, we get

$$\left( x_{\text{P}}' + x_{\text{Q}}' \right) \tan \vartheta - z_{\text{P}}' - z_{\text{Q}}' = 0 . \tag{V.6}$$

Taking the derivative of eq. (V.1) yields

$$\left( x_{\text{P}}' - x_{\text{Q}}' \right) \tan \vartheta - z_{\text{P}}' + z_{\text{Q}}' = \frac{L}{\cos \vartheta} . \tag{V.7}$$

The angular derivatives of the couples of variables $(x_{\text{P}}, z_{\text{P}})$ and $(x_{\text{Q}}, z_{\text{Q}})$ are related through the equation of the function $z = f(x)$. Using the chain rule for the derivatives of composite functions,

$$z_{\text{P}}' = \left. \frac{\text{d}f}{\text{d}x} \right|_{\text{P}} x_{\text{P}}' \quad ; \quad z_{\text{Q}}' = \left. \frac{\text{d}f}{\text{d}x} \right|_{\text{Q}} x_{\text{Q}}' , \tag{V.8}$$

from eqs. (V.6) and (V.7) we get a system of equations including only two unknown derivatives, $x_{\text{P}}'$ and $x_{\text{Q}}'$. Resolving this system yields

$$x_{\text{P}}' = \tfrac{1}{2} \frac{L}{\sin \vartheta - \left. \dfrac{\text{d}f}{\text{d}x} \right|_{\text{P}} \cos \vartheta} \quad ; \quad x_{\text{Q}}' = -\tfrac{1}{2} \frac{L}{\sin \vartheta - \left. \dfrac{\text{d}f}{\text{d}x} \right|_{\text{Q}} \cos \vartheta} . \tag{V.9}$$

From these expressions the derivatives of $x_{\text{O}}$ and $z_{\text{O}}$, respectively, $x_{\text{O}}'$ and $z_{\text{O}}'$ are straightforward.

By the way, the centre of flotation has a very particular role in this motion. Under a small variation of the angle of roll, $\vartheta \rightarrow \vartheta + \text{d}\vartheta$, its coordinates change as follows:

$$x_{\text{O}} \rightarrow x_{\text{O}} + x_{\text{O}}' \, \text{d}\vartheta \quad ; \quad z_{\text{O}} \rightarrow z_{\text{O}} + z_{\text{O}}' \, \text{d}\vartheta . \tag{V.10}$$

The initial and the new lines of flotation are described by, *cf.* eq. (V.1),

$$z_{\text{flotation}}(\vartheta) = z_O + (x - x_O)\tan\vartheta;$$

$$z_{\text{flotation}}(\vartheta + d\vartheta) = z_O + z_O{}'\,d\vartheta + (x - x_O - x_O{}'\,d\vartheta)\tan(\vartheta + d\vartheta), \qquad (V.11)$$

$$\approx z_O + (x - x_O)\tan\vartheta + \frac{x - x_O}{\cos^2\vartheta}d\vartheta$$

where we have made use of the fact that, according to eq. (V.6), $x_O{}'\tan\vartheta - z_O{}' = 0$. The centre of rotation (pivot point) of the line of flotation is the intersection of the initial and the new lines of flotation, therefore it verifies the system of equations (V.11) yielding

$$\begin{aligned} x_{\text{pivot}} &= x_O \\ z_{\text{pivot}} &= z_O \end{aligned}. \qquad (V.12)$$

Eqs. (V.11), (V.12) provide a simple but quite general proof of the Euler's theorem, see . Note that Euler himself had proved this theorem only in the vicinity of equilibrium. [11]

Taking the derivatives of eqs. (A.3), (A.4) with respect to $\vartheta$ and transforming the results by means of eqs. (V.9) yields

$$\begin{aligned} x_C{}' &= \tfrac{1}{12}\frac{L^3}{\mathcal{A}}\cos\vartheta \\ z_C{}' &= \tfrac{1}{12}\frac{L^3}{\mathcal{A}}\sin\vartheta \end{aligned}. \qquad (V.13)$$

Inserting $x_C{}'$ and $z_C{}'$ in the derivative of eq. (V.3) with respect to $\vartheta$, we get

$$E_h{}' = F_h\left[(x_C - x_O)\cos\vartheta + (z_C - z_O)\sin\vartheta\right]. \qquad (V.14)$$

Surprisingly, this equation looks as if in eq. (V.3) only the trigonometric functions were $\vartheta$-dependent, which is generally not the case.

Transforming to the Earth frame, *cf.* eq. (II.3), yields

$$E_h{}' = F_h X_C. \qquad (V.15)$$

Obviously, this result is consistent with the relation

$$dZ_C = -X_C\,d\vartheta \qquad (V.16)$$

arising from the fact that the rotation occurs about O and $X_O = 0$.

The rotational moment and the lever arm of the buoyancy force with respect to O are, *cf.* eq. (III.5),

$$M_{\mathrm{h}} = -F_{\mathrm{h}} X_{\mathrm{C}} \quad ; \quad a_{\mathrm{h}} = -X_{\mathrm{C}} . \tag{V.17}$$

In the position of stable equilibrium at $\vartheta = 0$ the *total* potential energy $E_{\mathrm{total}} = E_{\mathrm{h}} + E_{\mathrm{g}}$ of a floating body is minimal and the rotational moment of the couple $\boldsymbol{F}_{\mathrm{h}} + \boldsymbol{F}_{\mathrm{g}}$ vanishes. However, the hydrostatic energy at equilibrium does not necessarily take an extreme (minimal or maximal) value, so that in the general case the moment of the buoyancy force with respect to O, $\boldsymbol{M}_{\mathrm{h}}$ does not vanish. If at $\vartheta = 0$ $E_{\mathrm{h}}$ has a minimum, a deviation from equilibrium produces a restoring moment (of opposite sign to that of $\vartheta$). On the contrary, if at $\vartheta = 0$ $E_{\mathrm{h}}$ has a maximum, an overturning moment (of the same sign as that of $\vartheta$) is produced.

The elementary work of the buoyancy force in a small rotation about O is expressed as, *cf.* eq. (III.3),

$$\delta W_{\mathrm{h}} = F_{\mathrm{h}} \, \mathrm{d} Z_{\mathrm{A}} = -E_{\mathrm{h}}' \, \mathrm{d}\vartheta . \tag{V.18}$$

Comparing with eqs. (V.15), (V.16) yields

$$\mathrm{d} Z_{\mathrm{A}} = \mathrm{d} Z_{\mathrm{C}} , \tag{V.19}$$

*i.e.*, the vertical displacement of the point of application of $\boldsymbol{F}_{\mathrm{h}}$ coincides with that of C. Therefore, A is rigidly connected with C, and, as far as A and C are located on the same vertical line, one can consider that C *is* the "energetical" point of application of $\boldsymbol{F}_{\mathrm{h}}$ in rolling/pitching motion. This result is different from that for the translational motion, in which case one can consider that $\boldsymbol{F}_{\mathrm{h}}$ is applied at G, *cf.* eq. (IV.8).

## V. 2. Dynamical considerations

Taking the derivative of eq. (V.14) with respect to $\vartheta$, after a transformation yields the second derivative of $E_{\mathrm{h}}$:

$$E_{\mathrm{h}}'' = F_{\mathrm{h}} \left[ \tfrac{1}{12} \frac{L^3}{\mathscr{A}} - (x_{\mathrm{C}} - x_{\mathrm{O}}) \sin\vartheta + (z_{\mathrm{C}} - z_{\mathrm{O}}) \cos\vartheta - x_{\mathrm{O}}' \cos\vartheta - z_{\mathrm{O}}' \sin\vartheta \right] . \tag{V.20}$$

In the latter formula the terms in $x_{\mathrm{O}}'$ and $z_{\mathrm{O}}'$ describe the displacement of the point O with respect to the body frame. Meanwhile, eqs. (III.4) and (III.12), describing the angular derivatives of $E_{\mathrm{h}}$, refer to a coordinate system with origin at the centre of rotation. Therefore, eq. (V.20) should be considered in a coordinate system centered at O, in which case $x_{\mathrm{O}}'$ and $z_{\mathrm{O}}'$ vanish, and we get the "truncated" second derivative of $E_{\mathrm{h}}$:

$$E_h'' = F_h \left[ \tfrac{1}{12} \frac{L^3}{\mathscr{A}} - (x_C - x_O) \sin \vartheta + (z_C - z_O) \cos \vartheta \right]. \tag{V.21}$$

In the Earth frame $E_h''$ is further simplified to

$$E_h'' = F_h \left( \tfrac{1}{12} \frac{L^3}{\mathscr{A}} + Z_C \right), \tag{V.22}$$

yielding the *vertical* distance from the line of flotation to A, *cf.* eq. (III.12)₁:

$$d_{\text{flotation}}^A = Z_A = \frac{E_h''}{F_h} = \tfrac{1}{12} \frac{L^3}{\mathscr{A}} + Z_C. \tag{V.23}$$

On the other hand, the distance of C from the line of flotation is $-Z_C$ ($Z_C$ is negative in the system of coordinates chosen), therefore,

$$d_C^A = Z_A - Z_C = \tfrac{1}{12} \frac{L^3}{\mathscr{A}}. \tag{V.24}$$

In accordance with the Bouguer's theorem [10], this expression corresponds to the metacentric distance, *vide infra*. Therefore, one can conclude that the "dynamical" point of application of the buoyancy force *coincides with the metacentre*.

Note that Herder and Schwab [9] have carried out a similar calculation in the particular case of a rectangular immersed cross section in equilibrium position. However, in their equations (53, 55), corresponding to the above eqs. (V.22), (V.23), incorrectly appears the distance between G and M, in contradiction with the Euler's theorem. In fact, the latter distance should be calculated from the second derivative of the total potential energy, *cf.* eq. (III.13).

The existence of two different points of application of $\boldsymbol{F}_h$ (C and M) deduced from two different definitions ("energetical" and "dynamical") for the same type of motion may seem paradoxical. While these two points result in identical rotational moments, because the corresponding lever arm is the same, their locations on the vertical axis are different.

In order to understand the cause of this duality, the reader can consider the apparently trivial case of the point of application of the force of gravity $\boldsymbol{F}_g$ exerted on the floating body in rolling/pitching motion. Applying to $\boldsymbol{F}_g$ the same formalism, as before for $\boldsymbol{F}_h$, *cf.* eqs. (V.3), (V.4), (V.14), (V.20), (V.22), (III.12), (V.23), shows that both definitions of the point of application of $\boldsymbol{F}_g$ converge to the centre of gravity of the body.

*In fact, the different behaviour of $\boldsymbol{F}_g$ and $\boldsymbol{F}_h$ is due to the fact that the shape of the immersed volume does not remain constant in the course of rolling/pitching motion. The corresponding variation gives rise to the additional term in* *eq. (V.22) and ff., depending on the length of the line of flotation L and responsible for the duality of the point of application of $F_h$ .*

For a fully immersed body $L = 0$, so that M is merged with C, *cf.* eq. (V.24), and the above duality disappears.

# VI. Metacentre and ship stability

Actually, in the literature one can find three different definitions of the metacentre and the metacentric curve. We have chosen to designate them as the "geometrical" one (the evolute of loci of the centre of buoyancy), the Bouguer's one (relating the metacentric distance to the moment of inertia of the plane of flotation) and the "dynamical" one (suggested by the finding by Herder and Schwab that the "dynamically equivalent" point of application of the buoyancy force is the metacentre). [9]

Previously, the "geometrical" metacentre has been defined for any angle of heel whereas the Bouguer's metacentre was considered only in the vicinity of equilibrium and the "dynamical" metacentre only in the particular case of a "shoe-box" in equilibrium [9]. Besides, the existing demonstrations of the Bouguer's theorem are employing *ad hoc* geometrical constructions and are far from being rigorous, see Refs. [8 (pp. 45-46)], [21 (pp. 81-82)].

In this context, we would like to clarify the following two points:

(i)   Do the three definitions of the metacentre address one and the same point of the floating body and if yes, is this true for any angle of heel?
(ii)  What is the physical meaning of the metacentric curve for an arbitrary angle of heel?

To begin with, we express the relation between the metacentric curve $z_M(x_M)$ and the buoyancy curve $z_C(x_C)$ by means of the usual definition of the evolute of a curve as the locus of its centers of curvature: [25]

$$x_M = x_C - \frac{N}{D} z_C{}' \quad ; \quad z_M = z_C + \frac{N}{D} x_C{}' . \tag{VI.1}$$

where

$$N = x_C{}'^2 + z_C{}'^2 \quad ; \quad D = x_C{}' z_C{}'' - x_C{}'' z_C{}' . \tag{VI.2}$$

The expressions of $x_C{}''$ and $z_C{}''$ are calculated by taking the derivatives of eqs. (V.13) with respect to $\vartheta$ and inserting $x_P{}'$ and $x_Q{}'$ from eqs. (V.9), see Appendix I. Substituting these expressions in eq. (VI.2) yields

$$N = \tfrac{1}{144} \frac{L^6}{\mathscr{A}^2} \quad ; \quad D = \tfrac{1}{144} \frac{L^6}{\mathscr{A}^2}, \tag{VI.3}$$

so that eq. (VI.1) reduces to

$$x_{\mathrm{M}} = x_{\mathrm{C}} - z_{\mathrm{C}}{}' \quad ; \quad z_{\mathrm{M}} = z_{\mathrm{C}} + x_{\mathrm{C}}{}', \tag{VI.4}$$

hence, the "geometrical" metacentric distance is

$$d^{\mathrm{M}}_{\mathrm{C\ geometrical}} = \sqrt{\left(z_{\mathrm{M}} - z_{\mathrm{C}}\right)^2 + \left(x_{\mathrm{M}} - x_{\mathrm{C}}\right)^2} = \frac{z_{\mathrm{M}} - z_{\mathrm{C}}}{\cos\vartheta} = \tfrac{1}{12} \frac{L^3}{\mathscr{A}}. \tag{VI.5}$$

This distance is exactly the same as the distance from C to the "dynamical" point of application of the buoyancy force, *cf.* eq. (V.24), therefore the "dynamical" metacentre coincides with the "geometrical" one.

Now let us demonstrate the Bouguer's theorem for an arbitrary angle of heel. The statement of this theorem is as follows:

"***The metacentric distance is equal to the ratio of the moment of inertia I of the plane of flotation with respect to a horizontal axis and the immersed volume V of the displaced fluid***":

$$d^{\mathrm{M}}_{\mathrm{C\ Bouguer}} = \frac{I}{V}. \tag{VI.6}$$

As before, we consider the immersed cross-section of the floating body. In this case eq. (VI.6) becomes

$$d^{\mathrm{M}}_{\mathrm{C\ Bouguer}} = \frac{\mathscr{I}}{\mathscr{A}} \tag{VI.7}$$

where $\mathscr{I}$ is the moment of inertia of the line of flotation with respect to O,

$$\mathscr{I} = \int_{-\frac{1}{2}L}^{\frac{1}{2}L} \lambda^2 \, \mathrm{d}\lambda = \tfrac{1}{12} L^3 \,, \tag{VI.8}$$

hence,

$$\frac{\mathscr{I}}{\mathscr{A}} = \tfrac{1}{12} \frac{L^3}{\mathscr{A}}, \tag{VI.9}$$

which proves the theorem. As the right-hand side of this equation coincides with that of eq. (VI.5), the "Bouguer's" metacentre coincides with the "geometrical" one.

We conclude that for any angle of heel, all three above-mentioned definitions of the metacentre are equivalent:

$$d_{C \text{ dynamical}}^{M} = d_{C \text{ geometrical}}^{M} = d_{C \text{ Bouguer}}^{M} = \tfrac{1}{12} \frac{L^{3}}{\mathscr{A}}. \tag{VI.10}$$

Generalizing eqs. (VI.10) to the three-dimensional case, one should perform the integration along the longitudinal axis of the body to get the expression of the global rolling metacentric distance as

$$D_{M} = \tfrac{1}{12} \frac{1}{\mathscr{L}_{\max}} \int\limits_{0}^{\mathscr{L}_{\max}} \frac{\mathscr{L}(\vartheta, y)^{3}}{\mathscr{A}(y)} \mathrm{d}\,y \tag{VI.11}$$

where $\mathscr{L}_{\max}$ is the total length of the body. The pitching metacentric distance can be determined in a similar way.

From eqs. (V.15), (V.17) and (V.23) it is seen that the *metacentric height* (vertical distance from the line of flotation to the metacentre) is directly proportional to the angular derivative of the lever arm of the buoyancy force:

$$d_{\text{flotation}}^{M} = \frac{E_{h}{}''}{F_{h}} = -a_{h}{}'. \tag{VI.12}$$

Thus, the character (restoring or overturning) of the rotational moment of the buoyancy force depends on the sign (resp. positive or negative) of $d_{\text{flotation}}^{M}$, and the absolute value of the latter determines the rate of angular dependence of the former.

# VII. Conclusion

We have shown that the location of the point of application of the buoyancy force depends not only on the type of motion of the floating body (translation or rolling/pitching) but, in the latter case, also on the definition of this point. In translation this point remains fixed with respect to the centre of gravity of the body while in rolling/pitching it is subject to a duality. Namely, from the viewpoint of the work-energy relation it is fixed with respect to the centre of buoyancy while from the viewpoint of the rotational moment it is located at the metacentre. This peculiarity of the buoyancy force is due to the fact that, whereas the immersed body can still be considered as a rigid one, the shape of the displaced fluid does not remain constant in the course of the motion. Indeed, in the case of a completely immersed body, the shape of displaced fluid remains fixed, so that metacentre and the centre of buoyancy coincide.

The concept of non-uniqueness of the point of application of a resultant force seems quite unusual; nevertheless, as we have shown, this non-uniqueness is an inherent feature of the buoyancy force. While this finding is not expected to bring about changes in practical applications, it has a certain fundamental and educational interest for the mechanics of floating bodies. It would be interesting to find out whether some other physical forces do possess a similar non-uniqueness of the point of application.

Using the general approach based on the hydrostatic energy formalism, we have shown that the various definitions of the metacentre ("geometrical", "Bouguer's" and "dynamical"), in fact, concern one and the same distinct point of the immersed body. This finding holds (i) for any shape of the immersed body and (ii) not only in the vicinity of equilibrium but also for any angle of heel.

Besides, from the viewpoint of the rotational moment the metacentre proves to be the point of application of the buoyancy force in the rolling/pitching motion. These findings shed new light on the long-standing concept of the metacentre.

Another, more practical, implication of this study concerns the criterion of ship stability in relation to the location of the metacentre. Indeed, the metacentric height is proportional to the angular derivative of the lever arm of the buoyancy force. Thus, the character (restoring or overturning) of the rotational moment of the buoyancy force depends on the sign (resp., positive or negative) of the height of the metacentre above the line of flotation, and the absolute value of this height determines the rate of its angular dependence.

The model developed in the present study allows one to get analytical expressions for the location of the metacentre for the floating body of an arbitrary shape. Thus, it presents a certain interest for teaching and practice of naval mechanics and engineering, as well.

## Appendix I: Some expressions used in the main text

For the areas of fully and partially immersed pieces of the cross section of a floating body, respectively, type (i) and (ii), introduced in Section III, we get:

$$
\begin{aligned}
\mathscr{A}_{\mathrm{i}} &= \int_{x_{\mathrm{inf}}}^{x_{\mathrm{sup}}} \int_{f_{-}}^{f_{+}} \mathrm{d}z \, \mathrm{d}x = \int_{x_{\mathrm{inf}}}^{x_{\mathrm{sup}}} \left( f_{+} - f_{-} \right) \mathrm{d}x \\
\mathscr{A}_{\mathrm{ii}} &= \int_{x_{\mathrm{P}}}^{x_{\mathrm{Q}}} \int_{f}^{z_{\mathrm{flotation}}} \mathrm{d}z \, \mathrm{d}x = z_{\mathrm{Q}} L \cos\vartheta - \int_{x_{\mathrm{P}}}^{x_{\mathrm{Q}}} f \, \mathrm{d}x
\end{aligned} \tag{A.1}
$$

where $x_{\mathrm{inf}}$ and $x_{\mathrm{sup}}$ are the extreme abscissas of the fully immersed pieces, and $\mathrm{P}\left(x_{\mathrm{P}}, z_{\mathrm{P}}\right)$, $\mathrm{Q}\left(x_{\mathrm{Q}}, z_{\mathrm{Q}}\right)$ and $L$ are, respectively, endpoints and the length of the line of flotation for a body inclined through an angle $\vartheta$:

$$L = \frac{x_Q - x_P}{\cos\vartheta}. \tag{A.2}$$

The coordinates of the centre of buoyancy of the immersed cross-section are calculated by adapting to the two-dimensional case the general formulae eqs. (III.1), (III.9). The horizontal coordinates of C for the pieces of types (i) and (ii) are

$$x_{C_i} = \frac{1}{\mathcal{A}_2}\int_{x_{\inf}}^{x_{\sup}} x\int_{f_-}^{f_+} \mathrm{d}z\,\mathrm{d}x = \frac{1}{\mathcal{A}_2}\int_{x_{\inf}}^{x_{\sup}}\left(f_+ - f_-\right)x\,\mathrm{d}x$$
$$x_{C_{ii}} = \frac{1}{\mathcal{A}_1}\int_{x_P}^{x_Q} x\int_{f}^{z_{\text{flotation}}} \mathrm{d}z\,\mathrm{d}x = x_O + \frac{1}{\mathcal{A}_1}\left[\tfrac{1}{12}L^3\sin\vartheta\cos^2\vartheta + \int_{x_P}^{x_Q}\left(x_O - x\right)f\,\mathrm{d}x\right] \tag{A.3}$$

The corresponding vertical coordinates of C are

$$z_{C_i} = \frac{1}{\mathcal{A}_2}\int_{x_{\inf}}^{x_{\sup}}\int_{f_-}^{f_+} z\,\mathrm{d}z\,\mathrm{d}x = \frac{1}{2}\frac{1}{\mathcal{A}_2}\int_{x_{\inf}}^{x_{\sup}}\left(f_+^{\,2} - f_-^{\,2}\right)\mathrm{d}x$$
$$z_{C_{ii}} = \frac{1}{\mathcal{A}_1}\int_{x_P}^{x_Q}\int_{f}^{z_{\text{flotation}}} z\,\mathrm{d}z\,\mathrm{d}x = z_O + \frac{1}{\mathcal{A}_1}\left\{\tfrac{1}{2}L\left(\tfrac{1}{12}L^2\sin^2\vartheta - z_O^{\,2}\right)\cos\vartheta + \int_{x_P}^{x_Q}\left(z_O f - \tfrac{1}{2}f^2\right)\mathrm{d}x\right\}. \tag{A.4}$$

The expressions of the second derivatives of the coordinates of the centre of buoyancy used in <mark>Section VI</mark> are as follows:

$$x_C{}'' = -\tfrac{1}{8}\frac{L^3}{\mathcal{A}}\left(\frac{1}{\sin\vartheta - \dfrac{\mathrm{d}f}{\mathrm{d}x}\Big|_P \cos\vartheta} + \frac{1}{\sin\vartheta - \dfrac{\mathrm{d}f}{\mathrm{d}x}\Big|_Q \cos\vartheta} - \tfrac{4}{3}\sin\vartheta\right)$$
$$z_C{}'' = -\tfrac{1}{8}\frac{L^3}{\mathcal{A}\cos\vartheta}\left(\frac{\sin\vartheta}{\sin\vartheta - \dfrac{\mathrm{d}f}{\mathrm{d}x}\Big|_P \cos\vartheta} + \frac{\sin\vartheta}{\sin\vartheta - \dfrac{\mathrm{d}f}{\mathrm{d}x}\Big|_Q \cos\vartheta} + \tfrac{4}{3}\cos^2\vartheta - 2\right). \tag{A.5}$$

# Appendix II: Some special cases of immersed bodies

Here we consider the buoyancy, flotation and metacentric curves for several simple shapes of floating bodies. All subsequent results are readily obtained as particular cases of the general formulae derived above.

## AII. 1. Circular cylinder

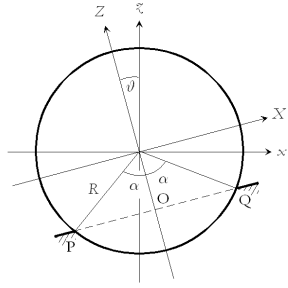

Consider the transversal cross-section of a floating long circular cylinder, its immersed part representing a circular segment of radius $R$ and centre angle $2\alpha$, see Figure 7. In this instance, the natural choice of the origin of coordinates, different from the previous one, is the centre of the circle. The immersed area is

$$\mathscr{A} = \tfrac{1}{2} R^2 \left( 2\alpha - \sin 2\alpha \right), \tag{A.6}$$

and the abscissas of the endpoints of the line of flotation are

$$x_{\mathrm{P}} = -R \sin\left(\alpha - \vartheta\right) \quad ; \quad x_{\mathrm{Q}} = R \sin\left(\alpha + \vartheta\right). \tag{A.7}$$

In the body frame both O and C describe circular arcs of equations

$$x_{\mathrm{O}}\left(\vartheta\right) = R \cos\alpha \sin\vartheta \quad ; \quad z_{\mathrm{O}}\left(\vartheta\right) = -R \cos\alpha \cos\vartheta \tag{A.8}$$

and, *cf.* eqs. (A.3), (A.4),

$$x_{\mathrm{C}}\left(\vartheta\right) = \tfrac{4}{3} R \frac{\sin^3\alpha}{2\alpha - \sin 2\alpha} \sin\vartheta \quad ; \quad z_{\mathrm{C}}\left(\vartheta\right) = -\tfrac{4}{3} R \frac{\sin^3\alpha}{2\alpha - \sin 2\alpha} \cos\vartheta, \tag{A.9}$$

see Figure 8. In particular, at equilibrium $x_{\mathrm{C}} = 0$ and $z_{\mathrm{C}}$ takes the values of $-R$, $-\tfrac{4}{3} R/\pi$ and $0$ respectively, for $\alpha = 0, \tfrac{1}{2}\pi$ and $\pi$. From eqs. (VI.4) it follows that the metacentric curve is reduced to a single point, the centre of the circle:

$$x_{\mathrm{M}} = 0 \quad ; \quad z_{\mathrm{M}} = 0, \tag{A.10}$$

and from eqs. (A.9) and (A.10) the metacentric distance is

$$d_{\mathrm{C}}^{\mathrm{M}} = \tfrac{4}{3} R \frac{\sin^3\alpha}{2\alpha - \sin 2\alpha}. \tag{A.11}$$

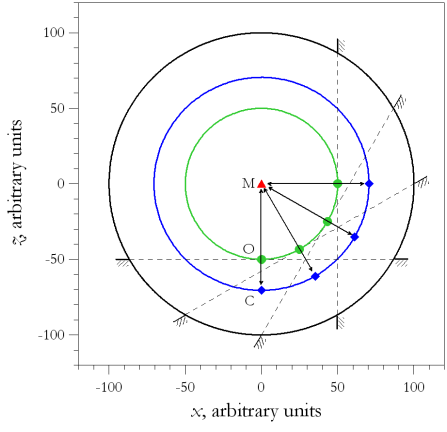

In accordance with eq. (V.3), the depth of C below the line of flotation is

$$d_C^{\text{flotation}} = R\left(\tfrac{4}{3}\frac{\sin^3\alpha}{2\alpha - \sin 2\alpha} - \cos\alpha\right). \tag{A.12}$$

Both distances and the hydrostatic energy $E_h = m_f\, g\, \mathscr{A}\, d_C^{\text{flotation}}$ are independent of $\vartheta$. As a consequence, the lever arm of the buoyancy force vanishes.

The floating equilibrium is stable if G is located below the centre of the circle (the axis of the cylinder).

## AII. 2. Rectangular box

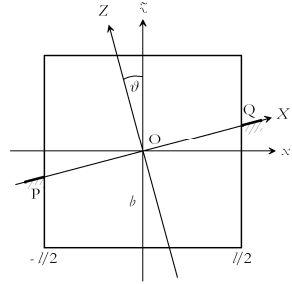

Figure 9. Floating square of side $l$. $b$ is the immersion depth at equilibrium.

Consider a floating body of rectangular cross-section (a box, see Figure 9) and denote $l$ its width and $b$ the immersion depth at equilibrium. The origin of coordinates is chosen at O. For definiteness we consider the case where two angles of the rectangle are immersed. Obviously, $\mathscr{A} = bl$, $x_P = -\tfrac{1}{2}l$ and $x_Q = \tfrac{1}{2}l$. The buoyancy and the metacentric curves as functions of the angle of heel are, cf. eqs. (A.3), (A.4), (VI.4),

$$x_C(\vartheta) = \tfrac{1}{12}\frac{l^2}{b}\tan\vartheta \quad;\quad z_C(\vartheta) = -\tfrac{1}{2}b + \tfrac{1}{24}\frac{l^2}{b}\tan^2\vartheta \tag{A.13}$$

and

$$x_{\mathrm{M}}(\vartheta) = -\tfrac{1}{12}\frac{l^2}{b}\tan^3\vartheta \quad ; \quad z_{\mathrm{M}}(\vartheta) = -\tfrac{1}{2}b + \tfrac{1}{24}\frac{l^2}{b}\left(\frac{3}{\cos^2\vartheta} - 1\right).$$ (A.14)

Both curves are illustrated in <mark>Figure</mark> 10. The metacentric curve is "V"-shaped, and for the metacentric distance we get

$$d_{\mathrm{C}}^{\mathrm{M}} = \tfrac{1}{12}\frac{l^2}{b}\frac{1}{\cos^3\vartheta}.$$ (A.15)

The depth of C below the line of flotation and the hydrostatic energy are, respectively, *cf.* eq. (V.3),

$$d_{\mathrm{C}}^{\mathrm{flotation}} = -Z_{\mathrm{C}} = \tfrac{1}{24}\frac{l^2}{b}\frac{\sin^2\vartheta}{\cos\vartheta} + \tfrac{1}{2}b\cos\vartheta .$$
$$E_{\mathrm{h}} = \mu_{\mathrm{f}}g_\star\mathcal{A}d_{\mathrm{C}}^{\mathrm{flotation}}$$ (A.16)

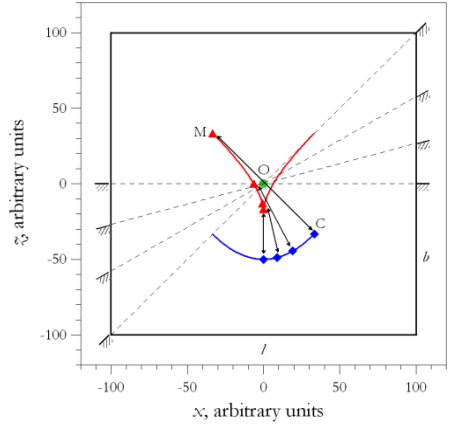

<mark>Figure</mark> 10. Buoyancy and metacentric curves for a floating square with $l = 200$ and $b = 100$. The diamonds and triangles indicate, respectively, the locations of C and M for $\vartheta = 0°, 15°, 30°$ and $45°$. The flotation curve is reduced to a point (circle).

The lever arm of the buoyancy force, in accordance with eq. (V.17), is

$$a_{\mathrm{h}} = \tfrac{1}{2}b\left[1 - \tfrac{1}{12}\frac{l^2}{b^2}\left(1 + \frac{1}{\cos^2\vartheta}\right)\right]\sin\vartheta .$$ (A.17)

The angular dependence of $a_{\mathrm{h}}$ is shown in <mark>Figure 11,</mark> left. For larger bodies $E_{\mathrm{h}}$ is minimal at equilibrium, $\vartheta = 0$ and increases with the angle of heel, *cf.* eq. (A.16); the sign of $a_{\mathrm{h}}$ is opposite to that of $\vartheta$, corresponding to a restoring moment. For narrower body an opposite behaviour is observed: at equilibrium $E_{\mathrm{h}}$ is maximal, so that $a_{\mathrm{h}}$ and $\vartheta$ have the same sign, corresponding to an overturning moment. In the vicinity of equilibrium the change in the sign of $a_{\mathrm{h}}$ takes place at a critical value $l/b = \sqrt{6}$ .

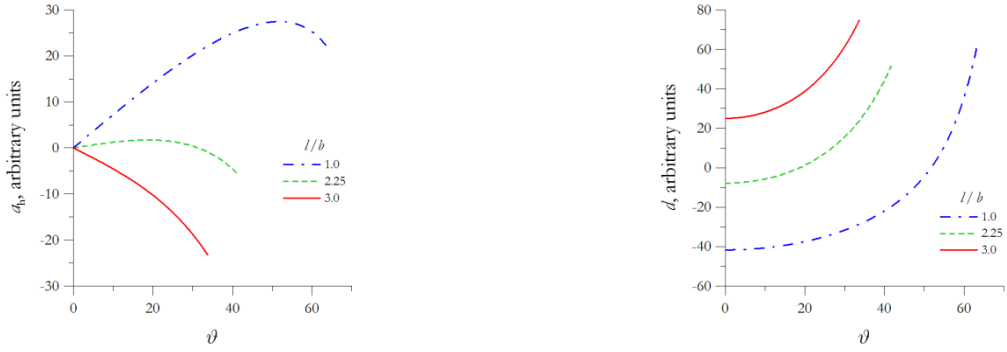

**Figure 11.** Left: lever arm of the buoyancy force for a floating rectangle of width $l$ and immersion depth $b$ vs. the angle of heel. Right: metacentric height. The $\vartheta$-values are limited by the requirement that of two angles of the rectangle be immersed. For $l/b = 1$ one always has an overturning moment, for $l/b = 2.25$ the moment is an overturning one for heel angles less than $\vartheta \approx 31°$ and becomes a restoring one at larger $\vartheta$. For $l/b = 3$ one always gets a restoring moment.

From eqs. (A.15) and (A.16), the metacentric height is, *cf.* eq. (VI.12),

$$d_{\text{flotation}}^{\text{M}} = -\tfrac{1}{2}\, b \cos \vartheta + \tfrac{1}{24}\, \frac{l^2}{b}\, \frac{2 - \sin^2 \vartheta \cos^2 \vartheta}{\cos^3 \vartheta}\,. \tag{A.18}$$

Figure 11, right shows the angular dependence of $d_{\text{flotation}}^{\text{M}}$.

The condition of stability of the vessel requires that at equilibrium ($\vartheta = 0$) G lie below M. From eq. (A.18), the limiting height of G with respect to the plane of flotation as a function of the $l/b$ ratio is

$$\left( \frac{Z_{\text{G}}}{b} \right)_{\text{lim}} = -\tfrac{1}{2} + \tfrac{1}{12} \left( \frac{l}{b} \right)^2 \tag{A.19}$$

The corresponding graph, shown in Figure 12, illustrates the amazing stability of a raft. Indeed, for $l/b \to \infty$, $Z_{\text{M}}/b \to \infty$ !

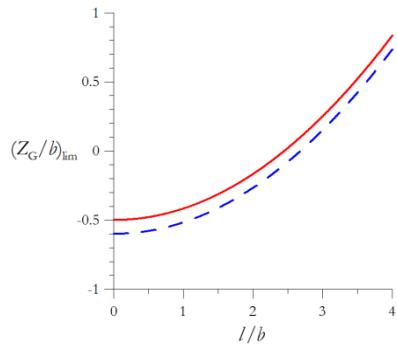

**Figure 12.** Limiting values of $Z_{\text{G}}$ for a rectangular box (full line) and a parabolic cylinder (*vide infra*, dashed line) for different widths $l$ (in relative units).

The special case of a "rolling rod" can be obtained from the above equations by assuming $l \to 0$. In this case, M coincides with C:

$$x_{\mathrm{M}} = x_{\mathrm{C}} = 0 \quad ; \quad z_{\mathrm{M}} = z_{\mathrm{C}} = -\tfrac{1}{2} b \, . \tag{A.20}$$

## AII. 3. Parabolic cylinder

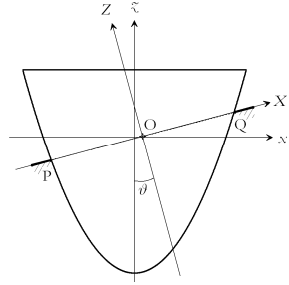

Consider a floating cylinder of cross section described by the parabola

$$z = 6 \frac{b}{l^2} x^2 - \tfrac{3}{2} b \, , \tag{A.21}$$

see Figure 13. The parameters in eq. (A.21) are chosen so as to facilitate a comparison with the rectangular box; indeed, $l$ is the length of the line of flotation at equilibrium, and the immersed area is $\mathscr{A} = bl$. The origin of coordinates is chosen at the centre of flotation at equilibrium, and the abscissas of the endpoints of the line of flotation are

$$x_{\mathrm{P}} = -\tfrac{1}{2} l + \tfrac{1}{12} \frac{l^2}{b} \tan \vartheta \quad ; \quad x_{\mathrm{Q}} = \tfrac{1}{2} l + \tfrac{1}{12} \frac{l^2}{b} \tan \vartheta \, . \tag{A.22}$$

In the body frame O and C describe parabolas of equations

$$x_{\mathrm{O}} = \tfrac{1}{12} \frac{l^2}{b} \tan \vartheta \quad ; \quad z_{\mathrm{O}} = \tfrac{1}{24} \frac{l^2}{b} \tan^2 \vartheta \tag{A.23}$$

and

$$x_{\mathrm{C}} = \tfrac{1}{12} \frac{l^2}{b} \tan \vartheta \quad ; \quad z_{\mathrm{C}} = -\tfrac{3}{5} b + \tfrac{1}{24} \frac{l^2}{b} \tan^2 \vartheta \tag{A.24}$$

(note that $x_{\mathrm{C}}$ and $x_{\mathrm{O}}$ coincide). The metacentric curve is given by, *cf.* eqs. (VI.4),

$$x_{\mathrm{M}} = -\tfrac{1}{12} \frac{l^2}{b} \tan^3 \vartheta \quad ; \quad z_{\mathrm{M}} = -\tfrac{3}{5} b + \tfrac{1}{24} \frac{l^2}{b} \left( \frac{3}{\cos^2 \vartheta} - 1 \right), \tag{A.25}$$

as in the case of floating box, it is "V"-shaped. The metacentric distance is, *cf.* eqs. (VI.5),

$$d_{\mathrm{C}}^{\mathrm{M}} = \tfrac{1}{12} \frac{l^2}{b} \frac{1}{\cos^3 \vartheta} \, . \tag{A.26}$$

The different characteristic curves are shown in Figure 14. There is much similarity between the cases of the rectangular box and the parabolic cylinder; indeed, in both cases the distance between the endpoints of the line of flotation remains constant, $x_Q - x_P = l$, and the metacentric distance is the same, *cf.* eqs. (A.15) and (A.26). Meanwhile, the centre of flotation of the box is fixed while that of the parabolic cylinder is a function of the angle of heel, *cf.* eq. (A.23). The expressions of the depth of C below the line of flotation and the hydrostatic energy for the parabolic cylinder are, *cf.* eq (V.3),

$$d_C^{\text{flotation}} = \tfrac{3}{5} b \cos\vartheta \quad ; \quad E_h = \tfrac{3}{5} \mu_f g b^2 l \cos\vartheta \,. \tag{A.27}$$

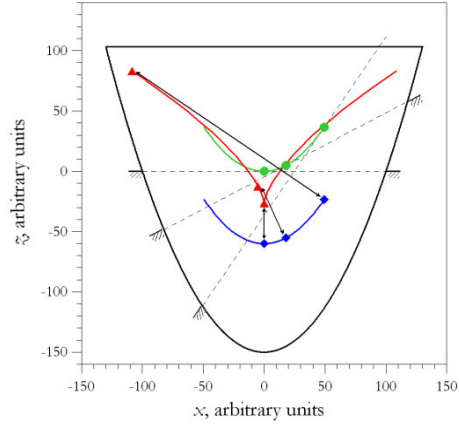

Figure 14. Buoyancy, flotation and metacentric curves for a floating parabola with $l = 200$ and $b = 100$. The diamonds, circles and triangles indicate, respectively, the locations of C, O and M for $\vartheta = 0°, 28°$ and $56°$.

As a difference from the rectangular box, for the parabolic cylinder the lever arm of the buoyancy force, *cf.* eq. (V.17), for $\vartheta > 0$ is always positive, resulting in an *overturning* moment

$$a_h = \tfrac{3}{5} b \sin\vartheta \,. \tag{A.28}$$

From eqs. (A.26) and (A.27), the metacentric height is, *cf.* eq. (VI.12),

$$d_{\text{flotation}}^M = \tfrac{1}{12} \frac{l^2}{b} \frac{1}{\cos^3\vartheta} - \tfrac{3}{5} b \cos\vartheta \,. \tag{A.29}$$

The limiting height of G with respect to the plane of flotation as a function of the $l/b$ ratio for $\vartheta = 0$,

$$\left( \frac{Z_G}{b} \right)_{\lim} = -\tfrac{3}{5} + \tfrac{1}{12} \left( \frac{l}{b} \right)^2 \,, \tag{A.30}$$

is illustrated in Figure 12, *vide supra*, in comparison with the rectangular box. In this case also $Z_M/b \to \infty$ for $l/b \to \infty$, so, G can be located very high above the plane of flotation, in spite of the above-mentioned tendency of the buoyancy force to overturn the body.

### AII. 4. Elliptic cylinder

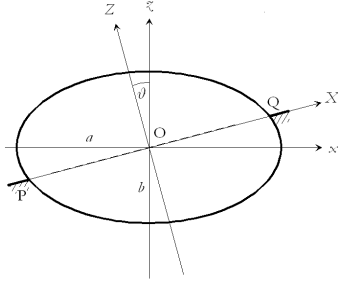

 15. Floating ellipse of semi-axes $a$ and $b$.

Consider an elliptic cylinder, see Figure 15, of cross-section described by

$$z = \pm \frac{b}{a} \sqrt{a^2 - x^2} \,. \tag{A.31}$$

where $a$ and $b$ are, respectively, the semi-major and semi-minor axes. As in the case of the circular cylinder, the origin of coordinates is chosen at the centre of the transversal cross-section. To simplify the formulae, we limit ourselves to the case of half-submerged cylinder, so that the line of flotation always passes through its centre and the area of its immersed half is $\mathscr{A} = \frac{1}{2}\pi ab$. Below we use the notation

$$R(\vartheta) = \sqrt{a^2 \sin^2 \vartheta + b^2 \cos^2 \vartheta} \tag{A.32}$$

($R(\vartheta)$ is the radius of the same ellipse turned through the angle $\vartheta + \frac{1}{2}\pi$).

In a position inclined through an angle $\vartheta$ the line of flotation meets the ellipse at the points

$$x_{P,Q} = \mp \frac{ab \cos \vartheta}{R(\vartheta)} . \tag{A.33}$$

The coordinates of C are

$$x_C = \tfrac{4}{3} \frac{a^2 \sin \vartheta}{\pi R(\vartheta)} \quad ; \quad z_C = -\tfrac{4}{3} \frac{b^2 \cos \vartheta}{\pi R(\vartheta)}, \tag{A.34}$$

and from eqs. (VI.4) one calculates the location of the metacentre:

$$x_M = \tfrac{4}{3} \frac{a^2 (a^2 - b^2) \sin^3 \vartheta}{\pi R^3(\vartheta)} \quad ; \quad z_M = \tfrac{4}{3} \frac{b^2 (a^2 - b^2) \cos^3 \vartheta}{\pi R^3(\vartheta)}. \tag{A.35}$$

The buoyancy and metacentric curves are shown in Figure 16 for rolling about two equilibrium states, respectively, with the major axis parallel and perpendicular to the line of flotation. In the first case the metacentric curve is "Λ"-shaped while in the second case it is "V"-shaped. From eqs. (A.34), (A.35) the metacentric distance is, *cf.* eq. (VI.5),

$$d_C^M = \frac{4}{3} \frac{a^2 b^2}{\pi R^3(\vartheta)}. \tag{A.36}$$

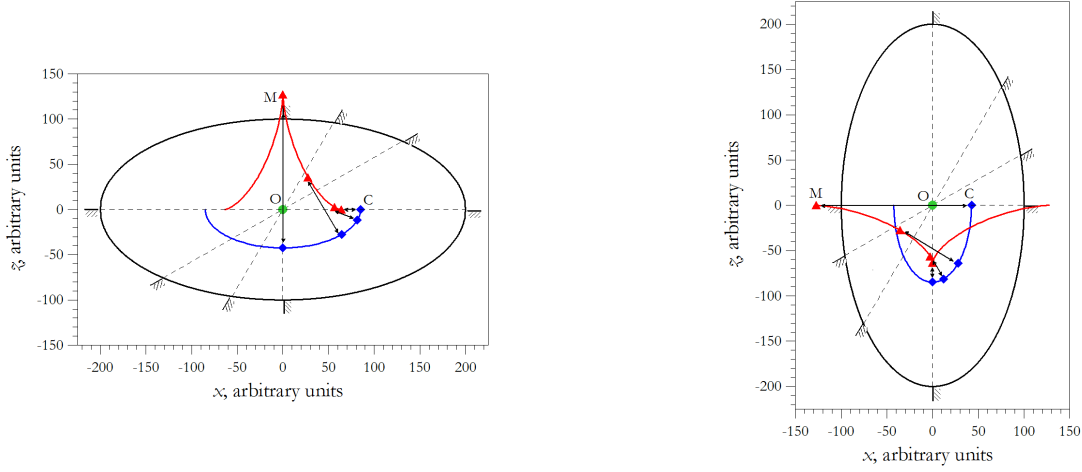

**Figure 16.** Buoyancy and metacentric curves for a floating ellipse with the major axis parallel (left) and perpendicular (right) to the line of flotation. The diamonds and triangles indicate, respectively, the locations of C and M for $\vartheta = 0°, 30°, 60°$ and $90°$. The flotation curve is reduced to a point (circle). The semi-major axis $a = 200$ and the semi-minor axis $b = 100$.

The depth of C below the line of flotation is, *cf.* eq. (V.3),

$$d_C^{\text{flotation}} = \frac{4}{3} \frac{R(\vartheta)}{\pi}, \tag{A.37}$$

so that for the hydrostatic energy we get

$$E_h = \frac{2}{3} \mu_f \, gab \, R(\vartheta). \tag{A.38}$$

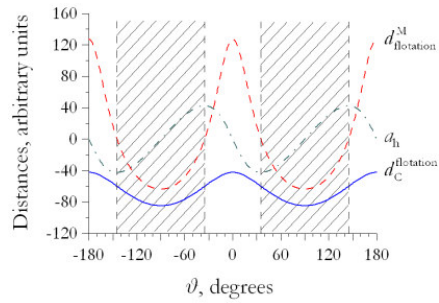

**Figure 17.** Depth of the centre of buoyancy (full line), metacentric height (dashed line), and lever arm of the buoyancy force (dashed-dotted line) for an elliptic cylinder with semi-major axis $a = 200$ and semi-minor axis $b = 100$. The non-shaded and shaded regions correspond, respectively, to restoring and overturning moment.

The lever arm of the buoyancy force, in accordance with eq. (V.17), is

$$a_h = -\frac{4}{3} \frac{a^2 - b^2}{\pi R(\vartheta)} \sin \vartheta \cos \vartheta. \tag{A.39}$$

From eqs. (A.36) and (A.37), metacentric height is, *cf.* eq. (VI.12),

$$d_{\text{flotation}}^{\text{M}} = \tfrac{4}{3}\frac{a^2 b^2 - R^4(\vartheta)}{\pi R^3(\vartheta)}. \tag{A.40}$$

Figure 17 shows the angular dependence of the locations of C and M with respect to the line of flotation as well as the lever arm of the buoyancy force with respect to O. At $\vartheta = 0, 180°$ the major axis is horizontal; the depth of C and the hydrostatic energy are minimal; the sign of $a_{\text{h}}$ is opposite to that of the deviation from equilibrium, corresponding to a restoring moment. At $\vartheta = 90°$ the major axis is vertical; the depth of C and the hydrostatic energy are maximal; the signs of $a_{\text{h}}$ and of the deviation from equilibrium, $\vartheta - 90°$ are the same, corresponding to an overturning moment. However, stable equilibrium corresponds to the minimum of the total potential energy and not to that of the hydrostatic energy only. For $\vartheta = 0$ the metacentric height is maximal and in the case shown in Figure 17 equilibrium remains stable for G located up to *ca.* 127 units *above* the line of flotation. On the other hand, for $\vartheta = \pm 90°$ the metacentric height is minimal and in order to preserve stable equilibrium, G should be located deeper than *ca.* 63.7 units below the line of flotation.